\documentclass[aps,prl,twocolumn,showpacs]{revtex4}
\usepackage{epsfig}

\begin{document}

\title{Measurement of Lagrangian velocity in fully developed turbulence}
\author{N. Mordant${}^{(1)}$, P. Metz${}^{(1)}$, O. Michel${}^{(2)}$,
J.-F. Pinton${}^{(1)}$}
\affiliation{${}^{(1)}$ CNRS \& Laboratoire de Physique, \'Ecole Normale
Sup\'erieure, 46 all\'ee d'Italie, F-69007 Lyon, France\\
${}^{(2)}$ Laboratoire d'Astrophysique, Universit\'e de Nice \\
Parc Valrose, F-06108, Nice, France}

\begin{abstract}
We have developed a new experimental technique to measure the
Lagrangian velocity of tracer particles in a turbulent flow, based
on ultrasonic Doppler tracking. This method yields a direct access
to the velocity of a single particle at a turbulent Reynolds
number $R_{\lambda} = 740$. Its dynamics is analyzed with two
decades of time resolution, below the Lagrangian correlation time.
We observe that the Lagrangian velocity spectrum has a Lorentzian
form $E^{L}(\omega)  = u_{\rm rms}^{2} T_{L} / (1 +
(T_{L}\omega)^{2} )$, in agreement with a Kolmogorov-like scaling
in the inertial range. The probability density function (PDF) of
the velocity time increments displays a change of shape from
quasi-Gaussian a integral time scale to stretched exponential
tails at the smallest time increments. This intermittency, when
measured from relative scaling exponents of structure functions,
is more pronounced than in the Eulerian framework.
\end{abstract}


\maketitle

Lagrangian characteristics of fluid motion are of fundamental
importance in the understanding of transport and mixing.  It is a
natural approach for reacting flows or pollutant contamination
problems to analyze the motion of individual fluid
particles~\cite{Pope}.  Another characteristic of mixing
flows is their high degree of turbulence.  For practical reasons, most
of the experimental work concerning high Reynolds number flows has
been obtained in the Eulerian framework.  Lagrangian measurements are
challenging because they involve the tracking of particle
trajectories: enough time resolution, both at small and large scales,
is required to describe the turbulent fluctuations.

Early Lagrangian information have been extracted from the dispersion
of particles, following Taylor's approach.  Recently numerical and
experimental studies have focused on resolving the motion of
individual fluid or tracer particles.  The emerging picture is as
follows.  The one-component velocity auto-correlation function is
quasi-exponential with a characteristic time of the order of the
energy injection scale~\cite{VirantDracos97,Sato,YeungPope89}.  The
velocity power spectrum is expected to have a scaling $E^{L}(\omega)
\propto \omega^{-2}$, as recently reported~\cite{Lien,Yeung2001} and
expected from a Kolmogorov similarity arguments.  In the same spirit,
the second order structure function should scale as $D^{L}_{2}(\tau) =
C_{0}\epsilon\tau$, where $\epsilon$ is the the power dissipation.
Measurements of atmospheric balloons~\cite{Hanna} have given $C_{0} =
4\pm2$ and a limit $C_{0} \rightarrow 7$ has been suggested in
stochastic models~\cite{Sawford}. Recent
experiments~\cite{VothBodenschatz98,PortaBodenschatz01} using high speed
optical techniques have shown that the statistics of the Lagrangian
acceleration are strongly non-Gaussian.

We have developed a new experimental method, based on sonar
techniques~\cite{MordantJASA}, in order to study in a laboratory
experiment the Lagrangian velocity across the inertial range of time
scales.  We obtain the first measurement of
single particle velocity for times up to the flow large scale turnover
time, at high Reynolds number.  In this Letter, we report the results
of this measurements and compare with previous observations and numerical
predictions.

Our technique
is  based on the principle of a continuous Doppler sonar.  A small
(2mm$\times$2mm) emitter continuously insonifies the flow with a
pure sine wave, at frequency $f_0 = 2.5$~MHz (in water). The
moving particle backscatters the ultrasound towards an array of
receiving transducers, with a Doppler frequency shift related to
the velocity of the particle: $2\pi \Delta f = \mathbf{q}.\mathbf{v}$.
The scattering wavevector $\mathbf{q}$ is equal to the difference
between the incident and scattered directions.  A numerical
demodulation of the time evolution of the Doppler shift gives the
component of the particle velocity along the scattering wavevector
$\mathbf{q}$.  It is performed using a high resolution parametric
method which relies on an Approximated Maximum Likelihood scheme
coupled with a generalized Kalman filter~\cite{MordantJASA}.  The
study reported here is made with a single array of transducers so that
only one Lagrangian velocity component is measured.

The turbulent flow is produced in the gap between two
counter-rotating discs~\cite{mor97}.  This setup has the advantage
to generate a strong turbulence in a compact region of space, with
no mean advection.  In this way, particles can be tracked during
times comparable to the large eddy turnover time.  Discs of radius
$R = 9.5$~cm are used to set water into motion inside a
cylindrical vessel of height $H=18$~cm. To ensure inertial
entrainment, the discs are fitted with 8 blades with height $h_{b}
= 5$~mm.  In the measurement reported here, the power input is
$\epsilon = 25$~W/kg.  It is measured on the experiment cooling
system, from the injection-dissipation balance. The integral
Reynolds number is $Re = R^2\Omega/\nu = 6.5 \; 10^{4}$, where
$\Omega$ is the rotation frequency of the discs (7.2~Hz), and $\nu
= 10^{-6}$~m$^{2}$/s is the kinematic viscosity of water.  A
conventional turbulent Reynolds number can be computed from the
measured $rms$ amplitude of velocity fluctuations ($u_{rms} =
0.98$~m/s) and an estimate of the Taylor microscale ($\lambda =
\sqrt{15\nu u_{rms}^{2} / \epsilon} = 0.88 \; $mm); we obtain
$R_\lambda = 740$.  This value is consistent with earlier studies
in the same geometry; it corresponds to the range of turbulent
Reynolds numbers where measurements of particle acceleration have
been reported~\cite{PortaBodenschatz01}.

The flow is seeded with a small number of neutrally buoyant
(density 1.06) polystyrene spheres with diameter $d=250 \; \mu$m .
It is expected that the particles follow the fluid motion up to
characteristic times of the order of the turbulence eddy turnover
time, at a scale corresponding to their diameter, i.e. $\tau_{\rm
min} \sim d/ u_{d} \sim \epsilon^{-1/3} d^{2/3}$, using standard
Kolmogorov phenomenology. For beads of diameter 250~$\mu$m, one
estimates $\tau_{\rm min} \sim 1.3$~ms. This value is within the
resolution of the demodulation algorithm whose cut-off frequency
is at 3~kHz. Note that the Kolmogorov dissipative time
($\tau_{\eta} = \sqrt{\nu/\epsilon} = 0.2$~ms) is smaller, so that
we do not expect to resolve the dissipative region. The
statistical quantities are calculated from $3\times 10^{6}$
velocity data points, taken at a sampling frequency equal to
6500~Hz. The acoustic measurement zone is in central region of the
flow, 10~cm thick in the axial direction and almost spanning the
cylinder cross-section. In this region the flow is a good
approximation to isotropic and homogeneous conditions: at all
points, the mean velocity is non zero, but equal to about
one tenth of its $rms$ value.

We first consider the Lagrangian velocity auto-correlation function:
\begin{equation}
    R^{L}(\tau) = \frac{\langle v(t) v(t+\tau) \rangle_{t}}{\langle
v^{2} \rangle} \; \; .
\end{equation}
We observe -- Fig.1a -- that it has a slow decrease which can be
modeled by an exponential function $\rho_{v}(\tau) \sim e^{-\tau /
T_{L}}$.  This expression defines an integral Lagrangian time scale
$T_{L} = 22$~ms.  For comparison, the period of rotation of the discs
is 140~ms and the sweeping period of the blades is 17~ms.  The
measured Lagrangian time scale thus appears as a time characteristic
of the energy injection.  The exponential reproduces extremely well
the variation of the auto-correlation function, from about $5
\tau_{\eta}$ at small scales to $4 T_{L}$.  These limits coincide with
the upper and lower resolution of the technique, so that we observe an
exponential decay over the entire range of our measurement.
However, as the variance of the acceleration must be
finite~\cite{PortaBodenschatz01} there has to be some lower cut-off to
this behavior, at times of order $\tau_{\eta}$.  These observations
extend and confirm previous numerical and experimental studies at
moderate Reynolds numbers~\cite{VirantDracos97,Sato,Yeung2001}.  Note
that the exponential decay of the Lagrangian velocity auto-correlation
is a key feature of stochastic models of dispersion since it appears
as a linear drift term in a Langevin model of particle
dynamics~\cite{Pope,Sawford}.

\begin{figure}[ht]
    \centering
    \epsfysize=11cm
    \epsfbox{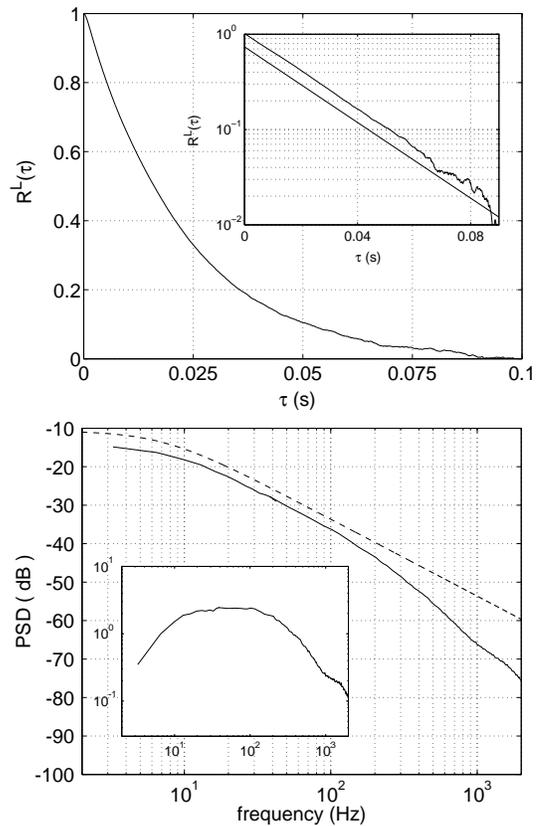}
    \caption{(a) Velocity autocorrelation function. A best exponential fit is
    $\rho^{L}_{v}(\tau) = 1.03 e^{-45.7 \tau}$. It is shown, slightly
    shifted for clarity, as the linear curve in the inset.  (b) corresponding
    power spectrum; the upper curve is the Lorentzian function
    calculated from the exponential fit of the auto-correlation
    function (shifted for clarity).}
\end{figure}

We show in Fig.1b the velocity power spectrum, computed both from the data
and as the Fourier transform of the exponential decay of the
auto-correlation function:
\begin{equation}
    E^{L}_{\rm fit}(\omega) = \frac{u_{rms}^{2} T_{L}}{1+
    (T_{L}\omega)^{2}} \ .
\end{equation}
We observe a clear range of power law scaling $E^{L}(\omega)
\propto \omega^{-2}$. This is in agreement with a Kolmogorov K41
picture in which the spectral density at a frequency $\omega$ is a
dimensional function of $\omega$ and $\epsilon$: $E^{L}(\omega)
\propto \epsilon \, \omega^{-2}$. To our knowledge, this is the
first time that it is directly observed at high Reynolds number
and in a laboratory experiment, although it has been reported in
oceanic studies~\cite{Lien} and in lower Reynolds number direct
numerical simulations~\cite{Yeung2001}. Departure from the
Kolmogorov behavior is observed at low frequencies in agreement
with the exponential decay of the auto-correlation. At high
frequencies, the spectrum deviates from the Lorentzian form due to
the particle response. Note in Fig.1b that the measurement is made
over a dynamical range of about 60~dB.

We now consider the second order structure function of the velocity
increment
\begin{equation}
    D_{2}^{L}(\tau) = \langle ( v(t+\tau) - v(t) )^{2} \rangle_{t} =
    \langle ( \Delta_{\tau} v )^{2} \rangle \ .
\end{equation}
We emphasize that these are time increments, and not space increments
as in the Eulerian studies.  The profile $D_{2}^{L}(\tau)$ is shown in
the inset of Fig.2.  It is linked to the auto-correlation by
$D_{2}^{L}(\tau) = 2 u_{\rm rms}^{2}\left( 1 - R^{L}(\tau) \right)$:
at small times one observes the trivial scaling $D_{2}^{L}(\tau)
\propto \tau^{2}$ and at large times $D_{2}^{L}(\tau)$ saturates at $2 u_{\rm
rms}^{2}$ (as $v(t)$ and $v(t+\tau)$ are
uncorrelated).

\begin{figure}[ht]
    \centering
    \epsfysize=5cm
    \epsfbox{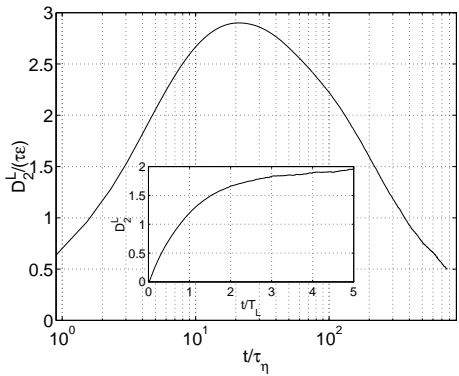}
    \caption{Second order structure function. Inset: profile
    $D_{2}^{L}(\tau)$ as a function of time, non-dimensionalized $T_{L}$.
    In the main figure the
    second order structure function is non-dimensionalized by the
    Kolmogorov scaling $\epsilon \tau$.}
\end{figure}

In between these two limits, one expects an inertial range of
scales with a Kolmogorov-like scaling
\begin{equation}
    D_{2}^{L}(\tau) = C_{0} \epsilon \tau \ ,
\end{equation}
where $C_{0}$ is a `universal' constant. Such a behavior is
consistent with dimensional analysis and with an $\omega^{-2}$
scaling range in the velocity power spectrum. Fig.2 shows
$D_{2}^{L}(\tau) / \epsilon \tau$;  a plateau with a constant
$C_{0}$ is not observed. Note that this also the case in Eulerian
measurements when the third order structure function is
represented in linear coordinates~\cite{mal98}. The function reaches
a maximum at $20 \tau_{\eta}$, for which $C_{0} \sim 2.9$. This
value is in agreement with the estimation $C_{0} = 4\pm 2$
in~\cite{Hanna} and in the range of values (between 3 and 7) used
in stochastic models for particle dispersion~\cite{Du}. 
 In our case there may also be a
bias at small times due to particle effects. However if we assume the exponential
fit for the velocity autocorrelation function to be valid down to the smallest scales, 
we obtain a value $C_0=3.5$ as an upper bound for the maximum of 
$D_{2}^{L}(\tau) / \epsilon \tau$.
In our set
of measurements between $R_{\lambda} = 100$ and $R_{\lambda} =
1100$, we have observed an increase of $C_{0}$ (defined in the
same way) from 0.5 to 4. We point out that in the absence of an
equivalent of the K\'arm\'an-Howarth relationship for the
Lagrangian time increments, a limit value of $C_{0}$ is not {\it a
priori} fixed. Dimensional analysis yields $D_{2}^{L}(\tau) =
C_{0}(Re) \epsilon \tau$ and similarity arguments give $C_{0}(Re)
\rightarrow {\rm const.}$ or  $C_{0}(Re) \rightarrow Re^{\alpha}$
in the limit of infinite Reynolds numbers.

To further describe the statistics of the Lagrangian velocity
fluctuations, we have analyzed the statistics of the
velocity increments $\Delta_{\tau} v$. Their PDF
$\Pi_{\tau}$ for $\tau$ covering the accessible range of time scales
is shown in Fig.3.

\begin{figure}[ht]
    \centering
    \epsfysize=5cm
    \epsfbox{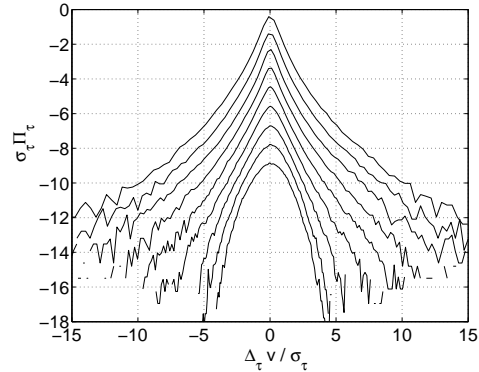}
    \caption{PDF $\sigma_{\tau}\Pi_{\tau}$ of the normalized increment
    $\Delta v_{\tau}/\sigma_{\tau}$. The curves are shifted for
    clarity. From top to bottom: $\tau = [0.15, 0.3, 0.6, 1.2, 2.5,
    5, 10, 20, 40]$~ms.}
\end{figure}

To emphasize the functional form, the velocity increments have been
normalized by their standard deviation so that all PDFs have unit
variance.  A first observation is that the PDFs are symmetric, in
agreement with the local symmetries this flow.  Another is that the
PDFs almost Gaussian at integral time scales and progressively develop
stretched exponential tails for small time increments.  At the
smallest increment, the stretched exponential shape is in agreement
with measurements of the PDF of Lagrangian acceleration at identical
Reynolds numbers~\cite{PortaBodenschatz01}.  In our case, the limit
form of the velocity increments PDF is not as wide as that of the
acceleration because the Kolmogorov scale is not resolved.  Note that
in regards of the evolution of the PDF, the intermittency is at least
as developed in the Lagrangian frame as it is in the Eulerian
one~\cite{Gagne}.

\begin{figure}[hb]
    \centering
    \epsfysize=5cm
    \epsfbox{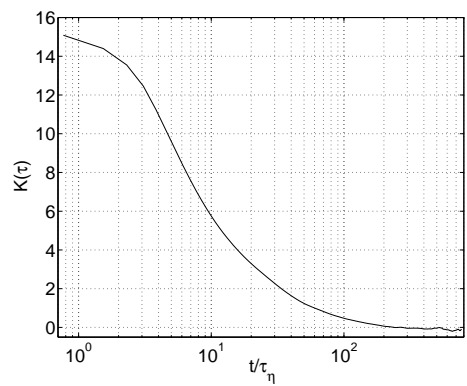}
    \caption{Evolution of the excess kurtosis factor $K(\tau) =
    \langle (\Delta_{\tau} v)^{4} \rangle  /  \langle (\Delta_{\tau} v)^{2}
    \rangle^{2} - 3$ for the PDFs of the time velocity increments.
    }
\end{figure}

The continuous evolution with scale can be quantified using the
flatness factor.  We show in Fig.4 the variation excess kurtosis
$K(\tau) = \langle (\Delta_{\tau} v)^{4} \rangle / \langle
(\Delta_{\tau} v)^{2} \rangle^{2} - 3$.  It is null at integral scale
as expected from the Gaussian shape of the PDF and increases steeply
at small scales.  Below about $5\tau_{\eta}$, the increase is limited
by the cut-off of the particle; an extrapolation of the trend
to $\tau_{\eta}$ yields $K(\tau_{\eta}) \sim 40$ in agreement with
acceleration measurements in~\cite{PortaBodenschatz01}.

\begin{figure}[h]
    \centering
    \epsfysize=5cm
    \epsfbox{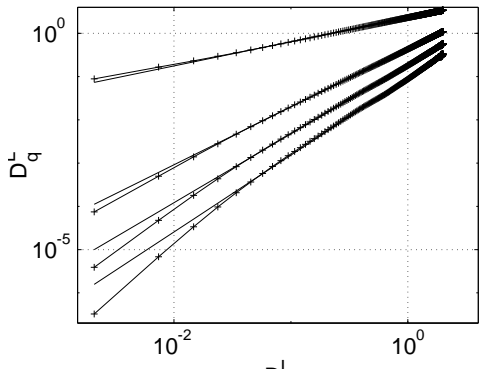}
    \caption{ESS plots of the structure function variation (in double
    log coordinates). The solid curves are best linear fits with
    slopes equal to $\xi^{L}_q = 0.56 \pm 0.01, 1.34 \pm 0.02, 1.56 \pm 0.06, 1.8 \pm 0.2$ for $p=1,3,4,5$
    from top to bottom.
    Coordinates in arbitrary units. }
\end{figure}

More generally, one can choose to describe the evolution of the PDFs
by the
behavior of their moments (or `structure functions') $D_{q}^{L}(\tau)
= \langle | \delta_{\tau} v |^{q} \rangle$.  Indeed, a consequence of
the change of shape of the PDFs with scale is that their moments, as
the flatness factor above, vary with scale.  Classically in the
Eulerian picture, one expects scaling in the inertial range,
$D_{q}^{E}(r) \propto r^{\zeta_{q}}$, at least in the limit of very
large Reynolds numbers.  At the finite Reynolds number where most
experiments are made, the lack of a true inertial range is usually compensated
by studying the relative scaling of the structure functions -- the ESS
ansatz~\cite{Benzi}.  We use the second order structure function as a
reference.  Indeed the dimensional estimation of $D_{2}^{L}$ (as that
of $D_{3}^{E}$) depends linearly on the increment and on the
dissipation.  Fig.5 shows that, as in the Eulerian frame, a relative
scaling is observed for the Lagrangian structure functions of orders
1 to 5, $D_{q}^{L}(\tau) \propto D_{2}^{L}(\tau)^{\xi_{q}}$.  We
observe that the relative exponents follow a sequence close to, but
more intermittent than the corresponding Eulerian quantity.  Indeed,
we obtain: $\xi^{L}_1/\xi^{L}_3 = 0.42$, $\xi^{L}_2/\xi^{L}_3 = 0.75$,
$\xi^{L}_4/\xi^{L}_3 = 1.17$, $\xi^{L}_5/\xi^{L}_3 = 1.28$ to be
compared to the commonly accepted Eulerian values~\cite{Arneodo96}
$\xi^{E}_1/\xi^{E}_3 = 0.36$, $\xi^{E}_2/\xi^{E}_3 = 0.70$,
$\xi^{E}_4/\xi^{E}_3 = 1.28$, $\xi^{E}_5/\xi^{E}_3 = 1.53$.

In conclusion, using a new experimental technique, we have obtained
a Lagrangian velocity measurement that covers the inertial range of
scales. Our results are consistent with
Kolmogorov-like dimensional predictions for second order
statistical quantities. At higher orders, the observed intermittency
is very strong. How the Lagrangian intermittency is related to the
statistical properties of the energy transfers is an open
question. From a dynamical point of view, the Navier-Stokes equation in
Lagrangian coordinates could be modeled using stochastic equations.
Work is currently underway to compare the dynamics of the Lagrangian
velocity to predictions of Langevin-like models. \\

{\bf acknowledgements:} We thank Bernard Castaing for interesting
discussions and Vermon Corporation for the design of the ultrasonic
transducers.  This work is supported by grant ACI No.2226 from the
French Minist\`ere de la Recherche.

\end{document}